\documentclass[pre,twocolumn,showpacs]{revtex4}

\usepackage{graphicx}
\usepackage{amsmath}

\begin{document}

\title{Hierarchical structure description of spatiotemporal chaos}

\author{Jian Liu$^1$, Zhen-Su She$^{1,2}\footnote{Correspondance: she@pku.edu.cn}$}
\affiliation{$^1$ State Key Laboratory for Turbulence and Complex
Systems \\
and Department of Mechanics and Engineering Science, Peking University, Beijing 100871, P.R. China\\
$^2$ Department of Mathematics, UCLA, Los Angeles, CA 90095, USA}

\author{Hongyu Guo, Liang Li, Qi Ouyang}
\affiliation{Department of Physics, Peking University, Beijing,
100871, P.R. China}

\date{\today}

\begin{abstract}
We develop a hierarchical structure (HS) analysis for quantitative
description of statistical states of spatially extended systems.
Examples discussed here include an experimental reaction-diffusion
system with Belousov-Zhabotinsky kinetics, the two-dimensional
complex Ginzburg-Landau equation and the modified FitzHugh-Nagumon
equation, which all show complex dynamics of spirals and defects.
We demonstrate that the spatial-temporal fluctuation fields in the
above mentioned systems all display the HS similarity property
originally proposed for the study of fully developed turbulence
[Z.-S. She and E. Leveque, Phys. Rev. Lett. {\bf 72}, 336 (1994)].
The derived values of a HS parameter $\beta$ from experimental and
numerical data in various physical regimes exhibit consistent
trends and characterize the degree of turbulence in the systems
near the transition, and the degree of heterogeneity of multiple
disorders far from the transition. It is suggested that the HS
analysis offers a useful quantitative description for the complex
dynamics of two-dimensional spatiotemporal patterns.
\end{abstract}

\pacs{05.45.-a; 82.40.Ck; 47.54.+r; 47.27.-i}

\vspace{5pt}

\maketitle

\section{Introduction}

Spatiotemporal chaos have been studied extensively in recent years
with different level of quantitative characterization
\cite{rev93}. It was discovered that for some spatially extended
systems such as the Rayleigh-B\'{e}nard (RB) convection
\cite{morris93,assenheimer94}, an ordered state of straight or
weakly curved rolls breaks down to a spatiotemporally chaotic
state consisting of elementary spiral structures which appear and
disappear in an irregular fashion. Spiral patterns driven far from
equilibrium have also been studied in chemical reaction-diffusion
systems such as Belousov-Zhabotinsky (BZ) reaction \cite{ouyang01}
and in biological systems such as cardiac muscle tissue
\cite{cardiac}. While an individual spiral is considered as a
self-organized topological defect, instabilities can lead to
so-called defect-mediated turbulence \cite{coullet89} that is
characterized by an exponential decay of correlations with length
and time. There have been very fruitful efforts in the
theoretical, experimental and numerical studies on the onset phase
of the instability which results in the breaking down of spiral
waves to the early transition to the disorder states (for a review
on the complex Ginzburg-Landau (CGL) system, see
Ref.~\cite{cgl02}). The study of the transition via instabilities
is important for problems such as the transition from spiral to
spiral turbulence in heart tissue which plays an essential role in
cardiac arrhythmia and fibrillation \cite{cardiac}.

Another question often raised is what quantitative measures are
useful when spatiotemporal patterns evolve, especially after
spatiotemporal chaos are developed \cite{cgl02,ch-science94}.
Attempt to identify defects and conduct statistical study of
defects in defect-mediated turbulence has yielded some interesting
results. In the discussion on the CGL system, it was argued
\cite{coullet89} that the number of defects can be a convenient
quantity to characterize spatiotemporal chaos. However, the
relation between the finite-time dimension and the number of
defects has not been fully established for systems of all sizes
and of all duration intervals, and thus the method of separation
has not appeared fully convincing \cite{cgl02}. In addition, when
spatiotemporal pattern become chaotic, the identification of the
basic structure (e.g. the defect) is not always possible, and its
statistics become dubious. In general, spirals act more like waves
than particles, but sometimes they may display spatially
intermittent features. Note that, although the RB convection, the
BZ reaction, and the CGL system all exhibit spiral wave behaviors
and defect structures, the instabilities leading to complex
spatiotemporal behaviors are of different nature. Moreover, local
pattern structures may also appear in various forms of spiral,
stripe, hexagon, or square. Nevertheless, transitions from ordered
to disordered states exist in all these systems, and a common
feature of these transitions is the generation of multiple scale
fluctuations. Therefore, simple phenomenology independent of the
details of the particular system is needed to reveal the essential
feature of spatiotemporal chaos ``buried in the wealth of
available data" \cite{ch-science94}. The present work propose such
a simple phenomenology but with a quantitative measure.

There are several other quantitative measures proposed for
characterizing disorder in pattern dynamics. Correlation length
was directly estimated in experiments \cite{ouyang91}, and was
also extracted from correlation functions \cite{cross95} or from
the width of so-called structure factors \cite{sf, note1}. Hu
\textit{et al.} \cite{hu95} proposed to use local wave numbers as
an order parameter to characterize experimental RB patterns.
Gunaratne \textit{et al.} \cite{gunaratne} suggested a disorder
function to describe disordered patterns close to locally striped
structures \cite{note2}. Egolf \textit{et al.} \cite{egolf98}
presented a simple algorithm for real-time quantitative analysis
of local structure by computing the local wave number in
disordered striped patterns. Newell and co-workers \cite{newell98}
used a wave-vector field determined by the wavelet transform to
study the behavior of phase-diffusion equations in the presence of
defects. All the methods mentioned above are effective to describe
irregular patterns with ``visually'' distinct local structures
such as local striped structure. When spatiotemporal chaos are
overwhelmingly developed, however, local striped structures can
not be defined and these measures can not be applied. This is the
case for spiral turbulence in highly turbulent regimes of BZ
experiments where either exact defect structures or characteristic
wave numbers are difficult to obtain. The present study gives a
different approach which may be applicable to much more disordered
patterns.

The BZ reaction system is a particularly interesting system for
the study of the spiral dynamics, where spirals can be supported
in excitable or oscillatory media \cite{rev93, ouyang01}. A recent
experimental study \cite{liu03a} has systematically identified
several regimes of spatial patterns resulted from two
instabilities, a Doppler instability and a long wavelength
instability. At various values of the chemical concentration in
reservoir A of malonic acid $[MA]^A$ and in reservoir B of
sulfuric acid $[H_2SO_4]^B$, the BZ reaction system displays a
variety of patterns including (see Fig.~1 which is reproduced from
Fig.~1 of Ref.~\cite{liu03a})
\begin{itemize}
  \item{simply rotating spiral ($S$);}
  \item{meandering spiral ($M$);}
  \item{chemical turbulence due to the Doppler instability ($D$);}
  \item{abnormal chemical turbulence ($A$);}
  \item{renascent stable spiral ($R$);}
  \item{convectively unstable spiral ($C_1$ and $C_2$);}
  \item{chemical turbulence due to the long wavelength instability ($T$)}
\end{itemize}
This rich set of spiral patterns is an ideal system for testing
theoretical methods for characterizing spatiotemporal chaos.

\begin{figure}[tbp]
  \includegraphics[width=8cm]{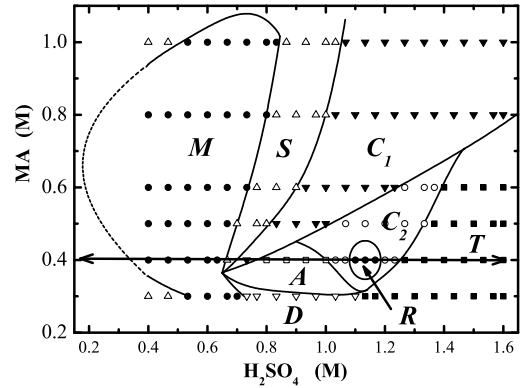}
  \caption{The phase diagram of the BZ experiment reproduced from
    \cite{liu03a}. The concentrations $[MA]^A$ and $[H_2SO_4]^B$ are
    the control parameters. The solid lines indicate the onsets of
    different instabilities. The dashed lines are the extrapolation of
    the solid lines. The heavy line with arrows at $[MA]^A=0.4M$
    indicate the case studied here.}
  \label{Fig:phase-diagram}
\end{figure}

We have proposed a hierarchical structure (HS) description of
spatiotemporal chaos \cite{liu03a, liu03b} based on an earlier HS
model of She and Leveque for hydrodynamic turbulence \cite{she94,
she98}. The starting point of the analysis is to construct a
multi-scale probability density function (PDF) description for the
fluctuation field of our interest, and then to develop a
phenomenological model for the description of PDFs over the entire
range of scales. This set of PDFs carries the probability
information for fluctuations at different scales (large versus
small) and at different intensities (tail events versus core
events). The scale enters as a parameter when the fluctuation
field is filtered for defining random variables, and the intensity
is parameterized by the order of the moments calculated with the
PDFs. Therefore, the dependence of the moments on the scale $\ell$
and on the order $p$ constitute a set of quantitative measures
about the properties of the field.

The HS model gives a special compact description of this set of
multi-scale and multi-intensity properties. The basic idea in the
HS model is that many statistically stationary multi-scale
fluctuation fields generated from nonlinear interactions between
many degrees of freedom have a self-organized property expressible
in terms of multifractal scaling whose multifractal dimension
function can be understood in terms of a similarity relation
between structures of increasing intensities of successive
moment-order $p$ \cite{she94}. This new similarity relation was
proposed for turbulence as a generalization of the Kolmogorov's
complete-scale-similarity \cite{she98}. Later, it was discovered
that the HS similarity is satisfied in a variety of nonlinear
systems and other complex systems, such as the RB convection
\cite{ching}, the Couette-Taylor flow \cite{she01}, flows in
rapidly rotating annulus \cite{she03c}, the climate variations
\cite{she02_climate}, the variation of nucleotide density along
DNA sequences \cite{she01_DNA}, the diffusion-limited aggregates
\cite{dla}, a fluctuating luminosity field of natural images
\cite{image} and several others \cite{passive}. More
interestingly, the previous studies show that the derived HS
parameters are closely related to global property of the turbulent
system. For example, in Couette-Taylor flow, one HS parameter
gives a signature of the breaking down of the Taylor vortex
\cite{she01}. It is thus hoped that the HS analysis would shed
light on the properties of a fluctuation field displaying
spatiotemporal chaos.

In \cite{liu03a,liu03b}, the HS analysis is briefly applied to
describe the spiral turbulence in the BZ experiment and in the
numerical solutions of the two-dimensional CGL equation. It was
shown that the HS similarity is accurately verified for various
experimental and numerical two-dimensional fluctuation fields
which include both the ordered spirals and spatiotemporal chaos.
In the study of the BZ experiment, it is discovered that the
measured HS parameter $\beta$ shows a transition when the system
undergoes a transition from one kind of spiral turbulence state to
another (e.g. from the state $D$ to the state $T$ in
Fig.~\ref{Fig:phase-diagram}). Those preliminary findings have
inspired the present detailed study of the nature of $\beta$ while
a more complex sequence of transitions are analyzed and more
numerical models are considered.

Specifically, we will examine more experimental data with
transitions between complex states. In
Fig.~\ref{Fig:phase-diagram}, at $[MA]^A=0.4$, the patterns vary
through $M$, $S$, $C_1$, $A$, $C_2$, $R$, and $T$. During the
transition, the amount of order (e.g. spiral structures), the
degree of heterogeneity (with the mixing of ordered and disordered
structures) and intermittency vary with parameters. We will show
that $\beta$ characterizes such transitions near and far from the
threshold in the experimental BZ system. We supplement this with a
further detailed study of the spiral dynamics in the CGL equation
\cite{cgl02} and in a two-component model equation called modified
FitzHugh-Nagumo (mFN) model \cite{barkley,bar} which has a number
of applications in biological systems. In all cases, we focus on
the validity of the HS description and the explanation of the
dynamics in terms of the degree of heterogeneity and of
intermittency that the value of $\beta$ indicates. We believe that
the relationship between the values of $\beta$ and the dynamics of
spatiotemporal patterns is valuable.

The paper is organized as follows. In Section II, we present
briefly the HS model with special emphasis on the method of the HS
similarity test ($\beta$-test) and the explanation of the meaning
of $\beta$. Section III is devoted to the analysis of a sequence
of complex transitions in the experimental BZ system. In Section
IV, we carry out a detailed study of the transition of the spiral
dynamics in the CGL equation and the mFN equation. Section V
offers a summary and some additional discussion.

\section{The HS description}

The HS model has been originally proposed by She and Leveque to
describe inertial-range multi-scale fluctuations in a turbulent
fluid. The key concept in the model is a hierarchy of moment
ratios. Let $\varepsilon_\ell$ denotes a certain characteristic
variable characterizing fluctuations of the physical system at the
length scale $\ell$. Define the $p$th-order moments
$S_p(\ell)=\langle|\varepsilon_\ell|^{p}\rangle$. The HS model
introduce a hierarchy of functions:
\begin{equation}
  F_p(\ell)=\frac{S_{p+1}(\ell)}{S_p(\ell)},
\end{equation}
each of which has the physical dimension of $\varepsilon_\ell$ and
hence describes a certain amplitude of the fluctuations. The
function $F_p(\ell)$ depends on the probability density functions
(PDFs) of $\varepsilon_\ell$. For a typical set of PDFs of a
turbulent field and of a spatiotemporal field as we study in this
work, the function $F_p(\ell)$ increases with $p$. In other words,
$F_p(\ell)$ at higher p is associated with higher intensity
fluctuation structures. We refer $F_p(\ell)$ to as the $p$th-order
HS function.

An intuitive idea in the HS model is that in a self-organized
dynamical steady state, there is a similarity in the dependence of
$F_p(\ell)$ on $\ell$ for different $p$'s ($0\le p\le\infty$). In
the case of scaling dependence on $\ell$, the simplest similarity
is that all $F_p(\ell)$ have the same scaling. This corresponds to
the Kolmogorov 1941 (K41) theory of turbulence in which the large
and small scale statistics are completely self-similar. For
example, when $\varepsilon_\ell$ represents the locally averaged
dissipation, then the K41 theory predicts that $F_p(\ell)\sim
F_0(\ell)=\langle\varepsilon_\ell\rangle \sim \ell^0$ (the mean
dissipation is equal to the average energy flux that is constant
independent of the scale). This complete self-similarity law has
been demonstrated invalid in experiments and numerical simulations
during the past decade, and the dissipation fluctuations have
anomalous scalings in $\ell$. This is so-called intermittency
effect \cite{sreenivasan91}.

She and Leveque \cite{she94} postulate that the most intense
structure $F_{\infty}(\ell)=\lim_{p\rightarrow \infty}F_{p}(\ell)$
plays a special role, and all other fluctuation structures of
finite $p$ obey a hierarchical similarity law, namely:
\begin{equation}
  \frac{F_{p+1}(\ell)}{F_{\infty}(\ell)}=A_{p}\left(\frac{F_{p}(\ell)}{F_{\infty}
   (\ell)}\right)^{\beta},
\label{hierarchical similarity}
\end{equation}
where $0 \le\beta\le 1$, $A_{p}$ are independent of $\ell$. The
term $F_{\infty}(\ell)$ can be eliminated by considering the ratio
\begin{equation}
  \frac{F_{p+1}(\ell)}{F_{p}(\ell)}=\frac{A_{p}}{
  A_{p-1}}\left(\frac{F_{p}(\ell)}{F_{p-1}(\ell)}\right)^\beta.
\label{beta-test}
\end{equation}
Both sides of Eq.~({\ref{beta-test}) can be computed from the
empirical PDFs (or histograms) of $\varepsilon_\ell$ calculated
from an experimental or numerical fluctuation field. The linearity
in the log-log plot of Eq.~(\ref{beta-test}) can be a direct test
of the validity of Eq.~(\ref{hierarchical similarity}). This will
be refer to as the HS similarity test, or the $\beta$-test
\cite{she01,she03a}. Figure~\ref{Fig:expt.beta-test} shows the
result of applying the $\beta$-test to the data of experimental BZ
reaction (for details, see below). The linear relation is
obviously verified, and the slope $\beta$ can be accurately
estimated.

\begin{figure}[tbp]
  \includegraphics[width=8cm]{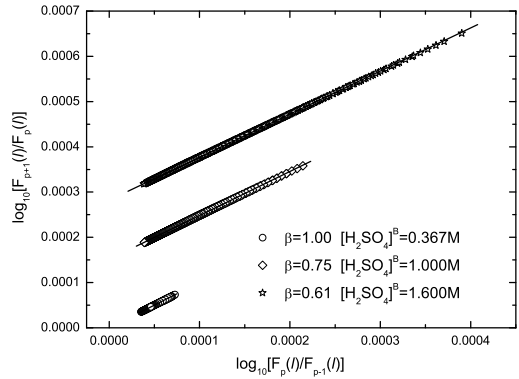}
  \caption{The HS similarity test for three patterns from the
    experimental BZ system: $[H_2SO_4]^B=0.367M$, $1.0M$, and $1.6M$.
    The other control parameter is fixed at $[MA]^A=0.4M$. A straight
    line indicates the validity of the HS similarity. The slope
    $\beta$ is estimated by a least square fitting. For clarity, the
    second and third set of data points are displaced vertically up by
    a suitable amount.}
  \label{Fig:expt.beta-test}
\end{figure}

The meaning of the parameter $\beta$ can be readily obtained from
the definition Eq.~(\ref{hierarchical similarity})
\cite{she98,she01}. It describes a kind of discrepancy between
$F_{p+1}$ and $F_p$. When $\beta$ approaches 1, high $p$ (intense)
structures and low $p$ (weak) structures are alike. This may be
realized at either extremely ordered states or completely
homogeneous disordered states. In the former, the structures of
various intensities are strongly correlated. Below, we will report
observations of this $\beta$ for ordered spiral states. The latter
situation corresponds to the hypothetical field postulated by the
K41 theory with complete self-similarity; unfortunately, it has
not yet been observed in real physical system. Another extreme
case is the limit $\beta \to 0$. This is the case in which
$F_\infty(\ell)$ stands out as the only kind of singular structure
which is responsible for the physical process. A mathematical
model displaying the dynamics of one-dimensional shocks, the
so-called Burgers equation, exhibits behavior close to this
description, where only shocks at isolated points are responsible
for the energy dissipation and all statistical moments are
determined by the discontinuities across the shocks.

The study of hydrodynamic turbulence reveals that in many
realistic turbulent systems, $0<\beta<1$, namely the most intense
fluctuations do not completely dominate and the fluctuation
structures of various intensities are not completely alike. The
fluctuation structures of various intensities are all responsible
for the physical process, but they are related by the hierarchical
similarity law Eq.~(\ref{hierarchical similarity}) which we
believe is a form of the self-organization of the system.

The results reported below on the experimental BZ system and
numerical spiral dynamics demonstrate that when the hierarchical
similarity holds, the parameter $\beta$ gives a quantitative
description of the degree of order/disorder and of
homogeneity/heterogeneity. For $\beta$ close to one, the system
appears to be orderly homogeneous and orderly self-organized; for
moderately smaller $\beta$ the system contains a mixture of
order/disorder and appears to be heterogeneous. The smallest
$\beta$ indicates the appearance of overwhelmingly disordered and
intermittent state with fully developed spatiotemporal chaos. The
details of these findings are reported below.

\section{Analysis of experimental BZ system}

The experimental BZ system analyzed here has been described in
detail in Ref.~\cite{liu03a}. An extensive phase diagram (see
Fig.~\ref{Fig:phase-diagram}) is established in the two
dimensional plane formed by the concentration $[H_2SO_4]^B$ versus
$[MA]^A$, in which various pattern dynamical states are identified
based (essentially) on visual inspection of experimental images or
luminosity fields. The HS analysis is performed on the total
variation field, $G(x,y)$, rather than directly on the luminosity
(image) field $f(x,y)$: which is proportional to the chemical
concentration:
\begin{equation}
  G(x,y)=\left[(\frac{\partial f}{\partial x})^2+(\frac{\partial f}{\partial y}%
    )^2\right]^{1/2}.
\label{Eq:g-define}
\end{equation}
Then, we construct a set of multi-scale variables to be the
coarse-grained local variation field, $\varepsilon_\ell$, similar
to the locally averaged energy dissipation in turbulence, which is
defined as below:
\begin{equation}
  \varepsilon_\ell={\frac 1{{\ell ^2}}}\int\limits_x^{x+\ell
  }\int\limits_y^{y+\ell }G(f(x',y'))dx'dy'.
\label{epsilon-eq}
\end{equation}
We examine the statistical properties by calculating the moments
of $\varepsilon_\ell$ as $\ell$ varying for increasing control
parameter $[H_2SO_4]^B$. The moments
$S_p(\ell)=\langle|\varepsilon_\ell|^{p}\rangle$ are calculated by
a space-average of $256\times 256$ pixels and by a time-average
over a number of images collected at different times. Furthermore,
the HS functions $F_p(\ell)$ are calculated and the ratio across
successive orders are evaluated and plotted, as in
Fig.~\ref{Fig:expt.beta-test}. It is readily seen that the
hierarchical similarity is satisfied and the HS parameter $\beta$
can be obtained (by a least square fitting) with a good accuracy.
Technically speaking, this completes the HS analysis of a
fluctuation field at a given control parameter.

\begin{figure}[tbp]
  \includegraphics[width=8cm]{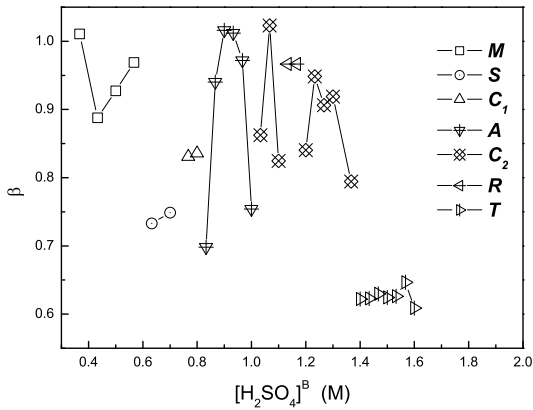}
  \caption{The HS similarity parameter $\beta$ as a function of the
    control parameter $[H_2SO_4]^B$ with fixed $[MA]^A=0.4M$. Symbols
    indicate the pattern regimes as found experimentally (see the text
    for detailed explanation).}
  \label{Fig:expt.all-beta}
\end{figure}

\begin{figure}[tbp]
  \includegraphics[width=8cm]{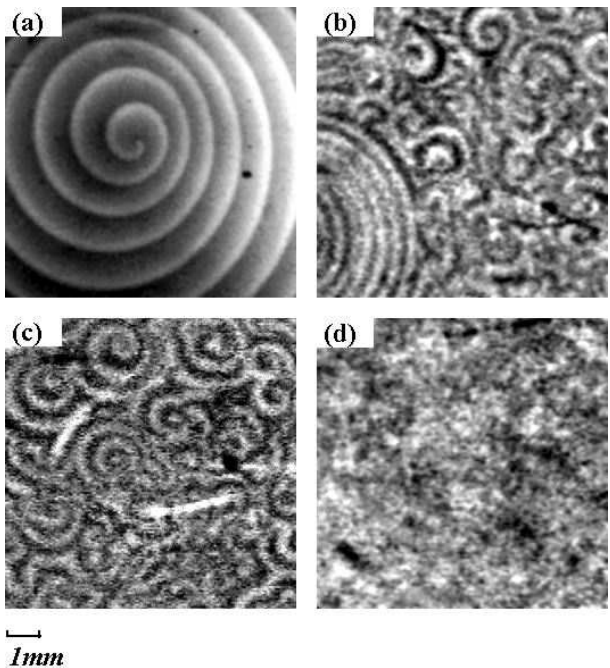}
  \caption{Optical images of typical experimental states obtained
    with different $[H_2SO_4]^B$: (a) an ordered state with $0.367M$, $\beta=1.0$;
    (b) a heterogeneous disordered state (coexistence of spiral waves
    and spatiotemporal chaos) with $0.833M$, $\beta=0.7$;
    (c) an abnormal turbulent state with $0.867M$, $\beta=0.94$;
    (d) a homogeneous intermittent disordered state with $1.4M$,
    $\beta=0.62$. The parameter $[MA]^A$ are fixed at $0.4M$.
    The region shown is $8.5 \times 8.5 mm^{2}$ out of a disk with $20 mm^{2}$ in diameter.}
  \label{Fig:expt.pattern}
\end{figure}

Detailed examinations of the values of $\beta$ at various control
parameters constitute a further step in the analysis. The
variation of $\beta$ for all patterns observed in the experiment
at each parameter value of $[H_2SO_4]^B$ is recorded in
Fig.~\ref{Fig:expt.all-beta}. The values of $\beta$ seem to
exhibit strong fluctuations, and it seems that there is no simple
correspondence between the value of $\beta$ and the regime. More
careful inspection reveals, however, that $\beta$ describes the
property of the pattern as expected from the theoretical
consideration above. We attempt to itemize our observations as
follows:

(a) For the ordered patterns, the values of $\beta$ measured are
close to one. In Fig.~\ref{Fig:expt.pattern}(a), we plot a typical
experimental image of the ordered spiral state ($M$) with
$\beta=1$. Theoretically speaking, $\beta=1$ implies that low and
high intensity structures are alike, which happens with an ordered
pattern. Note that some values of measured $\beta$ are not close
to one for ordered spiral states (e.g. the two $S$ states for
which $\beta\approx 0.75$ in Fig.~\ref{Fig:expt.all-beta}). One
possible explanation for the exception of the two $S$ states with
low $\beta$ is the existence of experimental noise (e.g., dark
spots in the experimental images) that introduces artificially
{\it sharp} structures in the field. Our studies show that such
``intrusion" of sharp objects lead to a decrease of $\beta$.

(b) Near the transition between one regime and another, we observe
a depletion of the value of $\beta$ compared to that measured for
the states far from the transition (see
Fig.~\ref{Fig:expt.all-beta}). We plot a typical experimental
image near the transition between $C_1$ and $A$ in
Fig.~\ref{Fig:expt.pattern}(b) with $\beta=0.7$. Theoretically
speaking, a lower value of $\beta$ indicates a higher
intermittency of the spatial patterns.
Fig.~\ref{Fig:expt.pattern}(b) indeed shows a mixture of ordered
spirals and disordered irregular spiral elements, which displays
clearly a strong heterogeneity. We conclude that one kind of
intermittency in spatially extended systems is related to
heterogeneous composition of patterns of different origins. In
addition, we conclude that a sharp variation of $\beta$ usually
indicates a transition of regimes. This finding is interesting for
defining statistical regimes. We believe that this is the most
interesting finding of this work.

(c) In fully developed spatiotemporally chaotic states, the
spatial patterns develop small-scale intermittent structures and
yield even smaller values of $\beta$ ($<0.65$). In
Fig.~\ref{Fig:expt.pattern}(d), we show a experimental image for
the chemical turbulent state $T$ with $\beta=0.62$. It is clear
that very small characteristic scales are developed in the
patterns which signify large fluctuations in the total variation
field (because of the spatial derivatives). This enhanced
intermittency appears to be stronger than noted in item (b).

(d) There are also other disordered states (e.g. $A$ and $C_2$)
for which $\beta$ is close to one. Close inspection shows that
these disordered states all encompass large-scale irregular spiral
elements, as shown in Fig~\ref{Fig:expt.pattern}(c) where
$\beta=0.94$. These large-scale patterns are quite homogeneous and
spatially ordered, which may yield a large value of $\beta$. Note
that the HS analysis conducted in this paper is a spatially
multiscaling analysis which describes only the property of
instantaneous spatial patterns. Our conclusion is that the states
in the regimes $A$ and $C_2$ involve only spatially ordered
patterns with actively temporal dynamics. They are distinguished
from the other chaotic states like $T$ and $D$.

The above analysis helps to understand the nature of the parameter
$\beta$ in the HS analysis of spatiotemporal chaos, and confirms
our previous finding that $\beta$'s variation reveals a transition
between two different spatiotemporally chaotic states. The spiral
turbulence state due to the Doppler instability has a
characteristic scale larger than that in the spiral turbulence
state due to the long wavelength instability. Both states are
homogeneously chaotic, but the later contains more intermittent
small-scale structures. The HS analysis is suitable to capture
such difference.

\section{Analysis of numerical spiral turbulence}

\begin{figure}[tbp]
  \includegraphics[width=8cm]{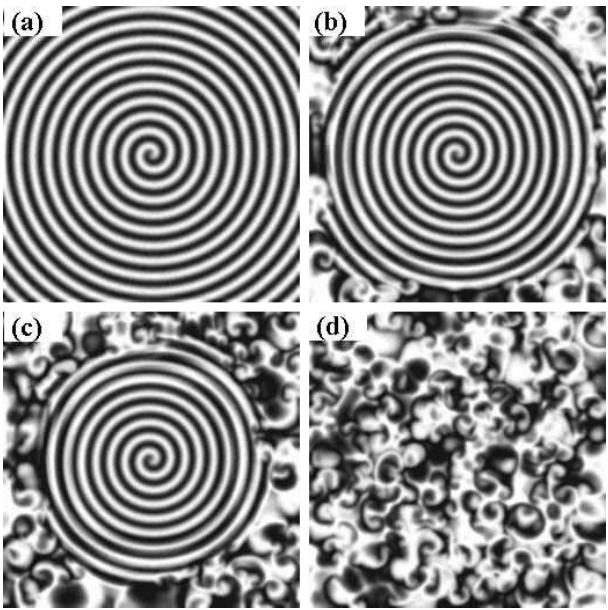}
  \caption{Snapshots of the field $f(x,y)$ from the CGL equation at
    four different parameter values of $c_3$. (a) stable spiral,
    $c_3=0.6$; (b) coexistence of spiral and spatiotemporal chaos,
    $c_3=0.74$; (c) coexistence of spiral and spatiotemporal chaos,
    $c_3=0.75$; (d) fully spatiotemporal chaos, $c_3=0.8$.}
  \label{Fig:CGL-c3-images}
\end{figure}

\begin{figure}[tbp]
  \includegraphics[width=8cm]{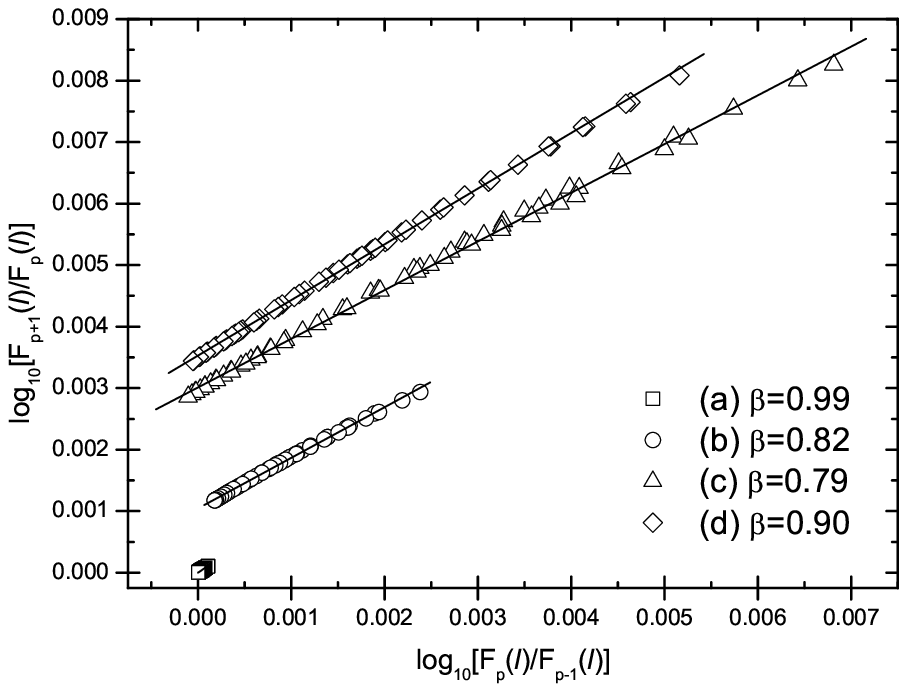}
  \caption{The HS similarity test for the four patterns in
    Fig.~\ref{Fig:CGL-c3-images} of the CGL equation. A straight line
    indicates the validity of the HS similarity. The slope $\beta$ is
    estimated by a least square fitting. For clarity, the second, the
    third and the fourth set of data points are displaced vertically
    up by a suitable amount.}
  \label{Fig:CGL-c3-beta-test}
\end{figure}

\begin{figure}[tbp]
  \includegraphics[width=8cm]{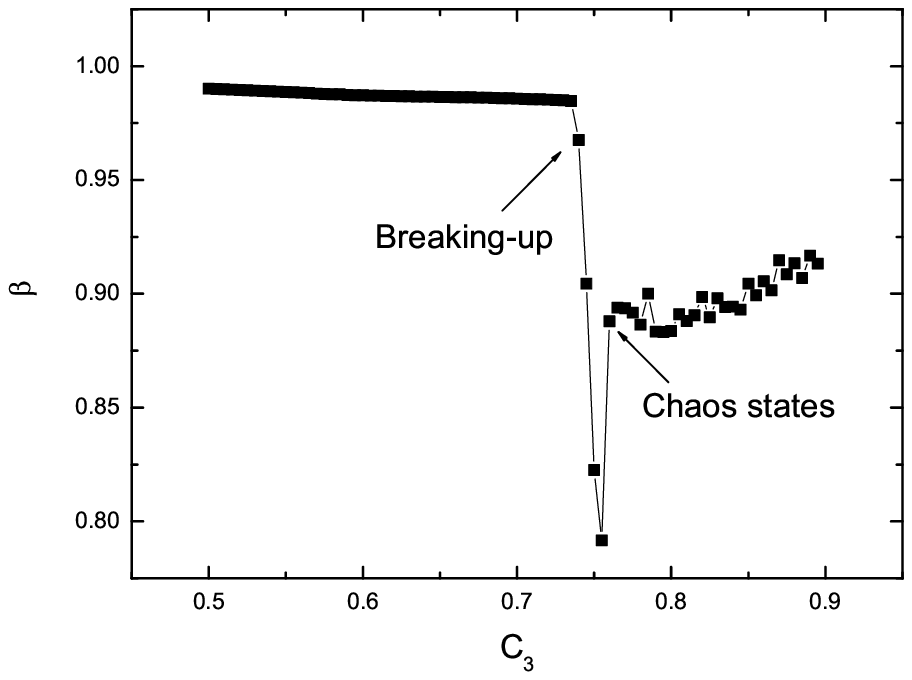}
  \caption{The HS similarity parameter $\beta$ as a function of the
    control parameter $c_3$ in the CGL equation. Note a clear
    transition from an ordered state to a spatiotemporally chaotic state.}
  \label{Fig:CGL-c3-all-beta}
\end{figure}

\begin{figure}[tbp]
  \includegraphics[width=8cm]{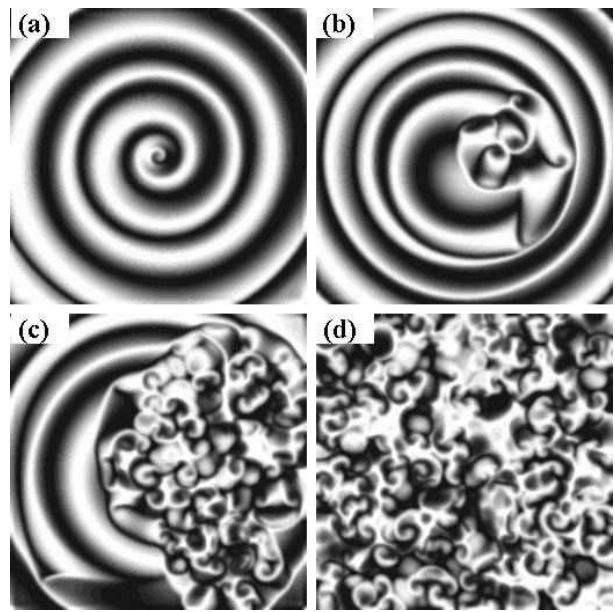}
  \caption{A sequence of evolving fields $f(x,y)$ from the CGL
    equation at different times. (a) $t=390$; (b) $t=500$; (c)
    $t=550$; (d) $t=770$.}
  \label{Fig:CGL-ev-images}
\end{figure}

\begin{figure}[tbp]
  \includegraphics[width=8cm]{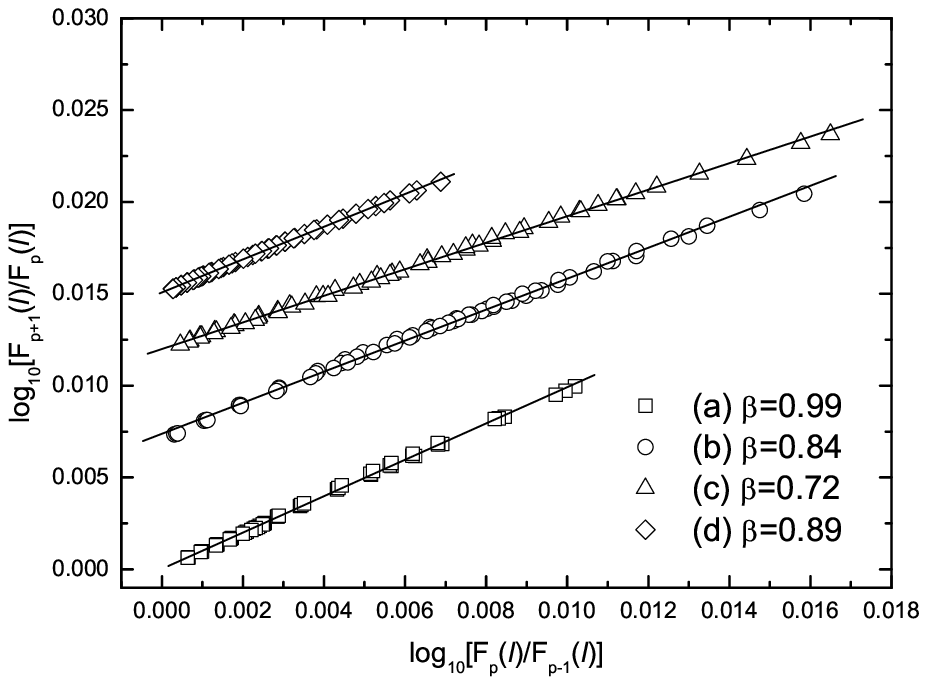}
  \caption{The HS similarity test for the four patterns in
    Fig.~\ref{Fig:CGL-ev-images} of the CGL equation. A straight line
    indicates the validity of the HS similarity. The slope $\beta$ is
    estimated by a least square fitting. For clarity, the second, the
    third and the fourth set of data points are displaced vertically
    up by a suitable amount.}
  \label{Fig:CGL-ev-beta-test}
\end{figure}

\begin{figure}[tbp]
  \includegraphics[width=8cm]{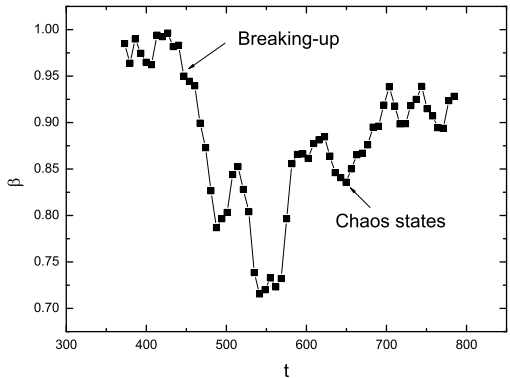}
  \caption{The HS similarity parameter $\beta$ as a function of time
    for a sequence of evolving patterns in the CGL equation. Note that
    $\beta$ acts as an order parameter revealing the transition from
    an ordered state to a spatiotemporally chaotic state.}
  \label{Fig:CGL-ev-all-beta}
\end{figure}

We now turn to the analysis of numerically simulated
two-dimensional spatiotemporal dynamics. First, in numerical
calculations, the interaction mechanism responsible for the
generation of spatial patterns is under good control, so that
spatial patterns are free of noises and perturbations that are
inevitable in experiments. For example, optical artifacts in the
experimental images pose great challenge to the HS analysis (in
the calculation of the total variation field) and thus compromise
the clarity of the finding. Secondly, the experimental states
analyzed above are all asymptotical states. We are also interested
in the characterization of transient states. In other words, we
are interested in testing whether the HS description can
characterize the property of a spatial pattern which undergoes
actively dynamical evolution. Numerical simulations provide a
detailed sequence of patterns under dynamic evolution. Finally,
the present analysis of numerically simulated field may inspire
further studies of a variety of two-dimensional partial
differential equations of pattern dynamics, e.g. the Navier-Stokes
equation for the vortex dynamics. Such empirical studies are
important to the finding of right order parameters of chaotic
pattern dynamics.

We will study two model equations for the pattern dynamics leading
to spatiotemporal chaos. The first equation is the two-dimensional
CGL equation:
\begin{equation}
  \frac{\partial A}{\partial t}=A+(1+ic_1)\triangle A
  -(1+ic_3)\left| A\right| ^2A,
\label{CGL-eq}
\end{equation}
where $A(\mathbf{r},t)$ is a complex function of time ($t$) and
space ($\mathbf{r}$), the real parameters $c_1$ and $c_3$ are
coefficients characterizing linear and nonlinear dispersion
respectively. Although the CGL equation is a normal form relevant
only near the threshold of a supercritical Hopf bifurcation, it
has become a popular model to study spatiotemporal chaos where
oscillations and waves are present \cite{rev93,cgl02}. Indeed, our
simulation gives rise to both simple spiral waves and chaotic
states which encompass many disordered spatiotemporal
oscillations.

The numerical study of Eq.~(\ref{CGL-eq}) is performed using an
Euler algorithm with no-flux boundary condition on a $256 \times
256$ square lattice and a time step of $\triangle t = 0.03$. The
initial conditions are chosen to be a simple spiral wave which is
the stable solution when $c_1=-1.40$ and $c_3=0.50$. We then keep
$c_1$ fixed and increase $c_3$ to explore possibly alternative
patterns. We choose to analyze the fluctuation properties of the
real part $f(x,y)$ of the complex field $A(x,y)= f(x,y)+ i
g(x,y)$. The details of the computation and the transition leading
to spatiotemporal chaos can be found in Ref.~\cite{liu03b}, where
we have established the fact that the $\beta$-test is satisfied in
both ordered spiral states and spatiotemporal chaos states. In the
present work, we report the result of a systematic HS analysis of
a sequence of patterns with varying parameters.

Four snapshots of the field $f(x,y)$ are shown in
Fig.~\ref{Fig:CGL-c3-images}, which contribute to a so-called
far-field breaking-up process. As $c_3$ increases, the convective
instability occurs and the spiral wave breaks up far from the tip.
In a range of intermediate values of $c_3$, the spiral wave
coexists with spatiotemporal chaos surrounding it. Finally, at
high values of $C_3$, the spiral wave totally breaks up and
spatiotemporal chaos dominate the whole box ($c_3>0.76$).

The HS analysis of the solutions of the CGL equation begins with
the construction of the total variation field. Let $G(x,y)$ be
defined as in Eq.~(\ref{Eq:g-define}) with the field $f(x,y)$. The
coarse-grained local variation field, $\varepsilon_\ell$, is
defined as in Eq.~(\ref{epsilon-eq}). The same procedure of the
$\beta$-test as explained in the previous section is now applied
to the data $\varepsilon_\ell$. As shown in
Fig.~\ref{Fig:CGL-c3-beta-test}, good linearity and hence the HS
similarity are obtained for all the states of both ordered spirals
and irregular spiral turbulence, consistent with the result
reported previously \cite{liu03b} and similar to that in
Fig.~\ref{Fig:expt.beta-test}. Note again that the values of
$\beta$ are close to one for ordered spiral states, but
significantly below one ($< 0.9$) when spatiotemporal chaos are
developed.

The variation of $\beta$ as a function of $c_3$ is shown in
Fig.~\ref{Fig:CGL-c3-all-beta}. All ordered spiral states have
$\beta$ very close to one, and the appearance of the irregular
patterns surrounding the central spiral leads to a smaller
$\beta$. It is interesting to note that the final chaotic states
at high $c_3$ have a $\beta\approx 0.9$ which is bigger than that
of the transitional states with a mixture of ordered spirals and
chaos. This fact is consistent with the previous finding in the
analysis of the experimental BZ system that the mixed states near
the transition are more heterogeneous and tend to have smaller
$\beta$.

Another interesting aspect of the numerical CGL dynamics is the
possibility to follow closely a spontaneous generation of spatial
disorder from an ordered state. A sequence of dynamically evolving
patterns are shown in Fig~\ref{Fig:CGL-ev-images}, where the
spiral core breaks up without meandering and then subsequently
develops to a complete spatiotemporal chaotic pattern. This fast
breaking-up process is phenomenologically similar to that of the
Doppler instability \cite{ouyang01,ouyang00}, but with the spiral
center fixed. The details of the explanation for this process can
be found in Ref.~\cite{liu03b}. Here, we show four typical
snapshots of the spatial pattern at successive times with fixed
control parameter $c_1=-1.40$ and $c_3=0.8$.

\begin{figure}[tbp]
  \includegraphics[width=8cm]{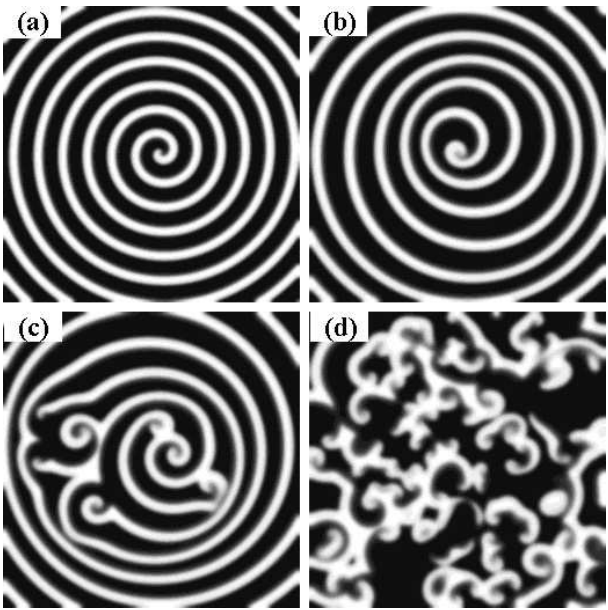}
  \caption{Snapshots of the field of the fast activator $u$ in the
    mFN equation. (a): simple spiral, $\sigma=0.0525$; (b):
    meandering spiral, $\sigma=0.07$; (c): outbreak of
    breaking-up spiral, $\sigma=0.0725$; (d): spatiotemporal
    chaos, $\sigma=0.075$. Other parameters are $a=0.84,
    b=0.07$.}
  \label{Fig:FN-images}
\end{figure}

Figure~\ref{Fig:CGL-ev-beta-test} reports the result of the
$\beta$-test for the corresponding four pattern states in
Fig.~\ref{Fig:CGL-ev-images}. Good HS similarity property is found
for all pattern states, including the single spiral
($\beta\approx1$), the intermediate state with generated defects
($\beta<0.9$) and the full developed turbulent state
($\beta\approx0.9$). Figure~\ref{Fig:CGL-ev-all-beta} shows the
variation of $\beta$ as a function of time during the core
breaking-up. The parameter $\beta$ begins with a value close to
one and undergoes a transition when disorder is developed. At
$t\approx 550$, $\beta$ reaches the minimum value when disorder is
mixed with order. Later, $\beta$ increases again because of the
homogeneity of the disorder. These facts are consistent with the
observations reported earlier. Once more, the HS parameter $\beta$
acts as an order parameter to the description of the evolution of
the CGL system from the ordered spiral to the disordered
spatiotemporal chaos.

The second model equation we have studied is a simple
activator-inhibitor model: the two-dimensional mFN model
\cite{barkley,bar}. While the CGL equation describes an
oscillatory dynamics (with two components representing the real
and imaginary parts of a complex variable field), the mFN model is
a two-component model of an excitable system in which a fast
activator ($u$) and a slow inhibitor ($v$) are in interaction:
\begin{subequations}
  \begin{equation}
    u_{t} = -\frac{1}{\sigma}u(u-1)\left[u-\frac{v+b}{a}
     \right]+\triangle u,
  \end{equation}
  \begin{equation}
    v_{t} = f(u)-v,
  \end{equation}
  \begin{eqnarray}
    f(u)  &=& \left\{
              \begin{array}{l@{\quad \quad}r}
                0 , & 0 \leq u<\frac{1}{3} \\
                1-6.75u(u-1)^{2}, & \frac{1}{3}\leq u \leq 1 \\
                1 , & u>1
             \end{array} \right.
  \end{eqnarray}
  \label{Eq:FN}
\end{subequations}
where, the form of $f(u)$ describes an inhibitor production only
above a threshold value of $u$, and the value of $\sigma$ related
to the time scales of both activator ($u$) and inhibitor ($v$)
measures the degree of excitability. A nice feature of the mFN
model is that the degree of the excitability of the medium can be
continuously modified by varying the parameters. It is known
\cite{bar} that fast activator dynamics ($0 < \sigma \ll 1$) and
suitable choice for other parameters ($a<1$, $b>0$) lead to
excitability. When one fixes the parameters $a$ and $b$ and
gradually increases $\sigma$, the excited spiral waves will loss
its stability, begin to meander, then break up at the tip, and
spread chaotic spiral elements outward. Finally, the system
develops a state of spatiotemporal chaos. This scenario is
illustrated in Fig.~\ref{Fig:FN-images}, which contains a sequence
of images of gray-scale picture of the activator $u$ obtained at
various times of a dynamic evolution.

\begin{figure}[tbp]
  \includegraphics[width=8cm]{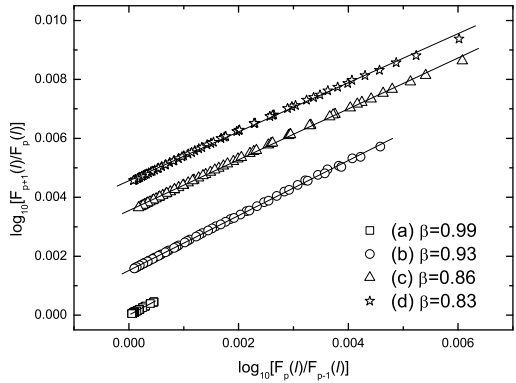}
  \caption{The HS similarity test for the four pattern states in
    Fig.~\ref{Fig:FN-images} of the mFN equation. A straight line
    indicates the validity of the HS similarity. The slope $\beta$ is
    estimated by a least square fitting. For clarity, the second, the
    third and the fourth set of data points are displaced vertically
    up by a suitable amount.}
  \label{Fig:FN-beta-test}
\end{figure}

\begin{figure}[tbp]
  \includegraphics[width=8cm]{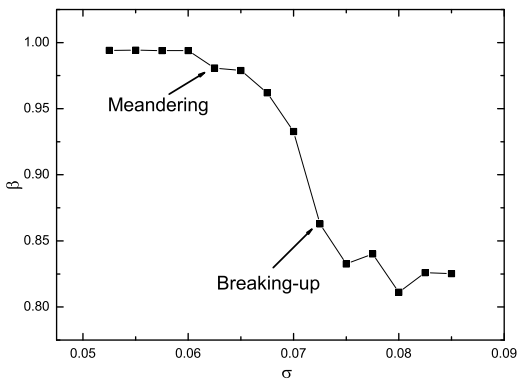}
  \caption{The HS similarity parameter $\beta$ as a function of the
    control parameter $\sigma$ in the mFN equation. Note a clear
    transition from an ordered state to a spatiotemporal chaotic state.}
  \label{Fig:FN-all-beta}
\end{figure}

The numerical calculation of Eq.~(\ref{Eq:FN}) reported above uses
an Euler algorithm with no-flux boundary condition on a $256
\times 256$ square lattice. The data presented in
Fig.~\ref{Fig:FN-images} are obtained with fixed parameters
$a=0.84$ and $b=0.07$, but with several different values of
$\sigma$. The simulation starts with a simple spiral initial
condition that is the stable solution of the system at smaller
values of $\sigma$ (e.g. $\sigma=0.50$). At $\sigma=0.725$, a
transition from the simple spiral to spatiotemporal chaos as
described earlier is fully realized.

To carry out the HS analysis, the total variation field $G(x,y)$
for the mFN model is calculated from the fast activator field
$u(x,y)$ as in Eq.~(\ref{Eq:g-define}). The coarse-grained local
variation field, $\varepsilon_\ell$, is defined as in
Eq.~(\ref{epsilon-eq}). We have applied the $\beta$-test for the
moments of $\varepsilon_\ell$, and the result is shown in
Fig.~\ref{Fig:FN-beta-test}. Clearly, the hierarchical similarity
is verified for all values of $\sigma$ and the values of $\beta$
are derived accurately in all cases. The variation of $\beta$ as a
function of $\sigma$ is presented in Fig.~\ref{Fig:FN-all-beta}.
The correspondence between the degrees of orders/disorder and the
values of $\beta$ is quite clear, as shown in the study of the CGL
equation. As the parameter $\sigma$ increases, the pattern
gradually loses the correlation and $\beta$ decreases. Detailed
examination shows that the tip of the spiral wave begins to
meander at $\sigma=0.625$, and completely breaks up at
$\sigma=0.725$. The break-up of the tip of the spiral marks the
transition from $\beta\approx 1$ to $\beta\approx 0.8$. The later
state corresponds to a fully developed spatiotemporal chaos.

The above numerical studies reveal that the CGL and mFN model
equations have less complicated patterns than those in the
experimental BZ system. This significantly reduces the complexity
of the matter, and renders the HS description of the dynamics easy
to interpret. It is worth noting that the numerical data are free
from complicated disturbance so that the values of $\beta$ show a
consistent trend of variation with less fluctuations, compared to
the results of Fig.~\ref{Fig:expt.all-beta}. We therefore find the
convincing evidence that the HS parameter $\beta$ is like an order
parameter that describes the transition from an ordered spiral
state to a homogeneous disordered spiral turbulence.

\section{Summary and discussion}

One of the most interesting problems in the study of the pattern
dynamics is to find a quantitative measure that describes the
global property of the pattern in both ordered states and
turbulent states. When spatiotemporal chaos prevail and the
effective number of degrees of freedom is large, many familiar
quantities in the study of dynamical systems and chaos become hard
to evaluate. On the other hands, traditional statistical measures
such as correlation length also become ineffective when the
pattern presents intermittent features. A new sound phenomenology
is needed to capture subtle correlations across multiple scales
and multiple fluctuation intensities.

We present here a sound phenomenology, the HS model, that gives a
concise description of a spatiotemporal field by its multifractal
scaling properties in terms of the HS parameter $\beta$. In this
work, we present detailed evidence, based on analyzing
experimental BZ system and numerical CGL and mFN model, that the
HS similarity holds for all the fields (ordered or disordered) and
the derived parameter $\beta$ describes the strength of the
correlation across scales and across intensities. For $\beta$
close to one, the pattern is strongly correlated or ordered; a
smaller $\beta$ implies a higher degree of heterogeneity and a
mixture of order and disorder; the smallest $\beta$ is related to
the intermittency associated with fully developed spatiotemporal
chaos. In particular, our analysis has identified two different
disordered states in the BZ experiment: the heterogeneous
disordered state with a mixture of spirals and irregular spiral
elements and the disordered and intermittent state of fully
developed spatiotemporal chaos. In those cases, $\beta$ provides a
useful characterization of the patterns in transition.

Based on the above results, we speculate that the HS similarity is
an emergent property of self-organized complex nonlinear
multi-scale systems in general. This property, like the maximum
entropy property of the thermal equilibrium state, may be
difficult to be derived directly from the first principle (such as
the Navier-Stokes equation for fluids or the Newtonian mechanical
law for gases). It is thus important to conduct more empirical
studies for proving/disproving the above conjecture, or for
assessing its limit of validity.

For the quantitative analysis of spatiotemporal chaos, we believe
that it is important to go beyond the simple statistics for
defects, wave numbers, or other local pattern properties, since
often they are hard to be determined when spatial patterns become
overwhelmingly irregular. The HS analysis is a global method that
takes into account the properties of the whole field, but
emphasizes at the same time subtle correlations and similarities
among structures at multiple scales and multiple intensities.
While the previous work \cite{liu03a, liu03b} establishes the
validity of the HS model for describing both experimental and
numerical spiral patterns, the present work gives detailed
evidence that the HS order parameter ($\beta$) characterizes the
degree of heterogeneity and intermittency of spatial patterns and
is particularly suitable for describing the evolution of patterns
from ordered states to fully developed spatiotemporally chaotic
states. This is demonstrated in the BZ, CGL and mFN systems. Our
approach may be easily extended to describe other complex
pattern-formation systems with different unit ``cells" than
spirals, such as strips, hexagons, squares, triangles, etc., and
this is one of the advantages of the HS analysis.

To make a concise description for all these statistical
properties, a phenomenology is required. The HS analysis is one of
the plausible phenomenological proposals that rest on empirical
observations ($\beta$-test). The fact that the HS similarity is
found to be valid for a wide class of nonlinear processes
indicates the possible universal mechanism/principle behind these
processes. It is not yet clear how this universality should be
expressed (e.g. in terms of a variational principle like the
maximum entropy and minimum entropy production), but we believe
that it deserves special attention in future studies.

\acknowledgments

We thank both anonymous referees for their helpful comments. We
have benefited from the research environment at the LTCS of Peking
University. The computation work is helped by the Turbulence
Simulation Center of the LTCS. We acknowledge the support by the
National Natural Science Foundation of China, No. 10225210, and by
the State Key Project for Basic Research No. 2000077305.


\begin{thebibliography}{99}
\bibitem{note2}
The disorder function calculates the local curvature of contour
lines and local characteristic wavenumber for patterns of local
stripped structures. For more complicated patterns, the
calculation of the disorder function is difficult. Detail comments
can be seen in Ref.~\cite{gunaratne}.
\end{thebibliography}
\end{document}